\documentclass[aps,prl,twocolumn,showpacs]{revtex4}
    \usepackage{graphicx,amssymb,amsthm}

\begin{document}

\title{Quantum Kalman Filtering and the Heisenberg
    Limit in Atomic Magnetometry}
\author{JM Geremia}
\email{jgeremia@Caltech.EDU}
\author{John K. Stockton}
\author{Andrew C. Doherty}
\author{Hideo Mabuchi}
\affiliation{Norman Bridge Laboratory of Physics, California
Institute of Technology, Pasadena, CA, 91125}
\date{\today}

\begin{abstract}
The shotnoise detection limit in current high-precision
magnetometry \cite{Romalis2003} is a manifestation of quantum
fluctuations that scale as $1/\sqrt{N}$ in an ensemble of $N$
atoms.  Here, we develop a procedure that combines continuous
measurement and quantum Kalman filtering \cite{Belavkin1999} to
surpass this conventional limit by exploiting conditional
spin-squeezing to achieve $1/N$ field sensitivity.  Our analysis
demonstrates the importance of optimal estimation for high
bandwidth precision magnetometry at the Heisenberg limit and also
identifies an approximate estimator based on linear regression.
\end{abstract}

\maketitle

\noindent Magnetometry is fundamentally a \textit{parameter
estimation} process because, like all fields, magnetism cannot be
directly observed.  Rather, the strength of a magnetic field must
be inferred from its influence on a probe such as an atomic spin
ensemble \cite{CohenTannoudji1969}.  In a canonical atomic
magnetometer, such an ensemble would be prepared into a coherent
spin state with its bulk magnetization polarized along the
$x$-axis, $\langle \hat{\mathbf{J}}(0) \rangle = (J,0,0)$ (such as
by optical pumping).  Then, a magnetic field along the $y$-axis
with magnitude, $B$, would induce the atomic Bloch vector,
$\langle \hat\mathbf{J}(t) \rangle$, to precess in the $xz$-plane
with frequency, $\omega_\mathrm{L} = \gamma B$. Thus, the magnetic
field could be estimated from the free induction decay of the
atomic magnetization by monitoring the $z$-component of the Bloch
vector, $\langle \hat{J}_\mathrm{z} (t) \rangle = J \exp(-t/T_2)
\sin(\omega_\mathrm{L} t)$, where $T_2$ is the transverse spin
coherence time.

In practice, current atomic magnetometers operate by continuously
pumping the atomic sample while a $\langle \hat{J}_\mathrm{z}
\rangle$-dependent optical property of the ensemble is monitored
\cite{Romalis2003,Budker2002,kim1998}.  Due to pumping, the atoms
are constantly re-polarized as they Larmor precess.  For small
fields (the relevant case when discussing detection limits), the
ensemble rapidly achieves an equilibrium that is nearly polarized
along the $x$-axis, but with a steady-state offset, $\langle
\hat{J}_\mathrm{z} \rangle_\mathrm{ss} \propto\gamma B J$. The
uncertainty in measuring $\hat{J}_\mathrm{z}$, is due to
projection noise \cite{Wineland1993}, $\langle \Delta
\hat{J}_\mathrm{z}^2 \rangle \equiv \langle \hat{J}_\mathrm{z}^2
\rangle - \langle \hat{J}_\mathrm{z} \rangle^2$, which has a value
of $J/2$ for a coherent spin state. Averaging a sequence of
independent measurements of $\langle \hat{J}_\mathrm{z}
\rangle_\mathrm{ss}$ with this variance leads to the conventional
shotnoise detection limit for a total measurement time of
$t_\mathrm{tot}$ \cite{Romalis2003,Budker2002},
\begin{equation} \label{Equation::ShotnoiseLimit}
    \delta B \simeq \frac{1}{\gamma \sqrt{J T_2
    t_\mathrm{tot}}} \, .
\end{equation}

Since $\langle \Delta \hat{J}_\mathrm{z}^2 \rangle$ sets an
intrinsic limit on the field sensitivity, reducing the projection
noise below its standard quantum limit would improve the
precision.  This naturally leads one to consider spin-squeezed
states \cite{Kitagawa1993} where uncertainty in $\langle
\hat{J}_\mathrm{z} \rangle$ is reduced by redistributing it into
the orthogonal spin component so that $\langle \Delta
\hat{J}_\mathrm{y}^2 \rangle > J/2$.  Since $\langle
\hat{J}_\mathrm{y} \rangle$ does not directly affect the field
estimation, spin-squeezing should enable one to surpass the
conventional shotnoise magnetometry limit.

An improved magnetometry protocol would ideally be implemented by
utilizing the conditional spin-squeezing that is automatically
generated by continuous observation of an atomic sample
\cite{Takahashi1999,Kuzmich2000,Thomsen2002}.  This dynamically
generated squeezing does not occur in steady-state (narrow-band)
magnetometers because of the continuous optical pumping. However,
it should be possible to enable sub-shotnoise magnetometry by
turning off the optical pumping once a coherent spin state has
been prepared followed by continuous observation of the atoms.

\begin{figure}[b*]
\hspace{-.3cm}
\includegraphics{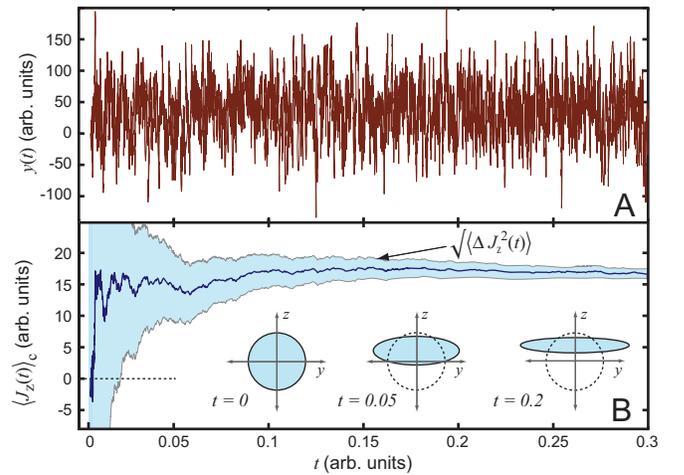} \caption{(A)
Simulatedsingle-shot atomic magnetometry photocurrent low pass
filtered at $F_\mathrm{c} = 2 \pi \sqrt{J}/t_\mathrm{tot}$.  (B)
Corresponding diffusion of the atomic Bloch vector as conditional
squeezing is produced by continuous QND observation.
\label{Figure::Photocurrent} }
\end{figure}

But the nature of conditional spin-squeezing gives rise to
potential complications that make it initially unclear how to
exploit the reduced uncertainty for improved magnetometry. Fig.\
\ref{Figure::Photocurrent} shows simulated data (generated
according to a quantum trajectory model described below) of a spin
ensemble under continuous measurement with no external field,
$B=0$.  As $\langle \Delta \hat{J}_\mathrm{z}^2 \rangle$ decreases
[shaded region in Fig.\ \ref{Figure::Photocurrent}(B)] with the
onset of spin-squeezing, there is no apparent change in the noise
of the associated $\langle \hat{J}_\mathrm{z}(t) \rangle$
measurement, $y(t)$ [Fig.\ \ref{Figure::Photocurrent}(A)], which
is due to constant optical shotnoise.

The dynamical generation of spin-squeezing starting from an
initial coherent state involves a stochastic transient at early
times.  As suggested by the error-ellipse diagrams of Fig.\
\ref{Figure::Photocurrent}(B), conditional evolution gradually
localizes the quantum spin state around a constant, but random,
value of $\langle \hat{J}_\mathrm{z} \rangle_\mathrm{c}$.  In an
ensemble of continuous measurement trajectories, this constant
value would be distributed with a variance of $J/2$ corresponding
to $\langle\Delta\hat{J}_\mathrm{z}^2 \rangle$ of the initial
coherent state.  Therefore, the mean value of $\langle
\hat{J}_\mathrm{z} \rangle_\mathrm{c}$ assumes a non-zero value
\textit{even in the absence of an applied magnetic field},
producing a stochastic offset in the photocurrent that must be
distinguished from Larmor precession in a magnetometry experiment.

Fortunately, with appropriate filtering, Larmor precession of the
spin state \textit{can} be distinguished from the projection noise
in such a way that the field estimation benefits from
spin-squeezing. In this Letter, we demonstrate that quantum
trajectory theory \cite{Carmichael1993,Wiseman1994} allows one to
construct a Kalman filter
\cite{Belavkin1999,Mabuchi1996,Verstraete2001,Stockton2003} that
optimally estimates the field magnitude from continuously observed
conditional atomic dynamics.  This filtering procedure enables
Heisenberg limited magnetometry despite the optical shotnoise and
the transient effects of spin state estimation. Furthermore, we
show that for time-invariant fields, our optimal strategy
approximately reduces to the simple and intuitive data analysis
procedure of linear regression which is a potentially simpler
experimental approach to sub-shotnoise magnetometry.

We propose a magnetometer in which the atomic ensemble undergoes a
continuous quantum non-demolition (QND) observation of
$\hat{J}_\mathrm{z}$. It has been shown that such a measurement
can be implemented by detecting $\hat{J}_\mathrm{z}$-dependent
changes in the phase of an off-resonant cavity mode coupled to the
atomic ensemble \cite{Thomsen2002} or by the Faraday rotation of a
far-detuned travelling mode \cite{Deutsch2003,Jessen2003} that
passes through the ensemble. In both cases the magnetometer
photocurrent is given by,
\begin{equation} \label{Equation::Photocurrent}
    y(t) \, dt = 2 \eta\sqrt{M} \langle \hat{J}_\mathrm{z}(t)
    \rangle_\mathrm{c} \, dt + \sqrt{\eta}dW(t) \,
\end{equation}
where $\langle \hat{J}_\mathrm{z}(t)\rangle_c$ is the conditional
expectation value of $\hat{J}_\mathrm{z}$, $\eta$ is the detector
efficiency and $M$ (in units of frequency) is an
implementation-dependent constant referred to as the
\textit{measurement strength}.  The optical shotnoise is reflected
by stochastic increments, $dW(t)$, that obey Gaussian white-noise
statistics, $E[dW]=0$ and $dW^2=dt$.

Conditional evolution of the atomic ensemble subjected to a
magnetic field along the $y$-axis and a QND measurement of
$\hat{J}_\mathrm{z}$ is described by the stochastic master
equation,
\begin{eqnarray}
    d\hat{\rho}_c(t) & = & -i\gamma B[\hat{J}_\mathrm{z},
            \hat{\rho}_c]dt +
            M \mathcal{D}[\hat{J}_\mathrm{z}]\hat{\rho}_c dt
            \label{Equation::SME}\\
      & &   + \sqrt{M\eta} \mathcal{H}[\hat{J}_\mathrm{z}]
            \hat{\rho}_c dW(t) \nonumber
\end{eqnarray}
where $\hat{\rho}_c(t)$ is the reduced atomic density operator
conditioned on the measurement record \cite{Wiseman1994}. The
super-operators, $\mathcal{D}$ and $\mathcal{H}$, are given by,
$\mathcal{D}[r]\hat{\rho} = r \hat{\rho} r^\dagger - (r^\dagger r
\hat{\rho} + \hat{\rho} r^\dagger r)/2$ and
$\mathcal{H}[r]\hat{\rho} = r \hat{\rho} + \hat{\rho} r^\dagger -
\mathrm{tr}[ (r + r^\dagger)\hat{\rho}] \hat{\rho}$, and the
initial condition is an optically-pumped coherent spin state along
the $x$-axis, $\hat{\rho}(0) = |J\rangle_x \langle J|$.

Each term in  Eq.\ (\ref{Equation::SME}) has a physical
implication for magnetometry.  First, the Hamiltonian, $H(B) =
\gamma B \hat{J}_y$, generates the desired Larmor precession
signal used to detect the magnetic field.  The second term
reflects measurement induced atomic decoherence that results from
coupling the ensemble to the optical shotnoise on the probe laser.
As a result the length of the Bloch vector decays over time,
$J(t)=J \exp(-Mt/2)$, and $M$ can be related to a bound on the
transverse spin relaxation, $T_2 \le 2M^{-1}$.

The significance of the third term in Eq.\ (\ref{Equation::SME})
is best seen after employing an approximation that holds for a
large net magnetization and small field ($\omega_L t \ll 1$). For
an ensemble polarized along the $x$-axis, quantum fluctuations in
$\hat{J}_x$ are at least second order and the operator,
$\hat{J}_x$, is well-approximated by the length of the Bloch
vector, $\hat{J}_x \rightarrow J$. Physically, this assumption
capitalizes on the large value of $J$ to treat the Bloch sphere as
a locally flat phase space.  This approximation is extremely good
for both coherent and squeezed states with $J \gg 1$. In a
Gaussian approximation, the first and second moments of
$\hat{J}_\mathrm{z}$ are sufficient to completely characterize the
atomic state. Therefore, equations of motion for the mean and
variance,
\begin{eqnarray}
    d \langle \hat{J}_\mathrm{z} \rangle_c & = & \gamma B
    J e^{-Mt/2} \, dt
        + 2 \sqrt{M\eta}
        \langle \Delta \hat{J}_\mathrm{z}^2 \rangle \, dW(t) \,\,\,\,
        \label{Equation::MeanJz} \\
    d \langle \Delta \hat{J}_\mathrm{z}^2 \rangle & = & -4 M \eta
    \langle \Delta \hat{J}_\mathrm{z}^2 \rangle^2 \, dt
    \label{Equation::DeltaJz2} \vspace{-.2cm}
\end{eqnarray}
provide a closed representation of the magnetometer's conditional
quantum dynamics (in the $\omega_L t \ll 1$ and $t \alt M^{-1}$
limits).  The physical significance of Eq.\
(\ref{Equation::MeanJz}) is that the atomic Bloch vector
experiences two types of motion: deterministic Larmor precession
and stochastic diffusion. Equation (\ref{Equation::DeltaJz2})
reflects the deterministic reduction of $\langle \Delta
\hat{J}_\mathrm{z}^2 \rangle$ as the atomic state is localized by
the observation process, i.e., conditional spin-squeezing.

Equations (\ref{Equation::MeanJz}) and (\ref{Equation::DeltaJz2})
can be used to implement an optimal estimation procedure that
capitalizes on squeezing without mistaking measurement-induced
Bloch vector rotations for true Larmor precession. Since the
atomic dynamics are stochastic, the estimator must be described
probabilistically---we desire a conditional probability
distribution, $p \left(B | \Xi_{[0,t]} \right)$, which measures
the likelihood that the field has magnitude $B$ given the
measurement record, $\Xi$, defined in terms of the photocurrent,
$d\,\Xi_{\,t} \equiv y(t)\,dt / (2\eta\sqrt{M})$.  The estimated
magnitude, $\tilde{B}$, and its uncertainty, $\Delta \tilde{B}^2$
are obtained from the moments,
\begin{eqnarray}
    \tilde{B} & = & \int B p \left(B | \Xi_{[0,t]} \right) \,
    dB\\
    \Delta \tilde{B}^2  & = &
        \int (B-\tilde{B})^2 p \left(B | \Xi_{[0,t]} \right) \,
        dB
\end{eqnarray}
of the conditional distribution, $p \left(B | \Xi_{[0,t]}\right)$.

Constructing a maximum-likelihood estimator is accomplished by
defining an update rule that iteratively improves $p \left(B
|\Xi_{[0,t]} \right)$ as the measurement record is acquired. Prior
knowledge of the distribution of magnetic field values is encoded
in $p\left(B|\Xi_0\right)$, which may be assigned infinite
variance in order to assure an unbiased estimate. Optimality
requires that the conditional probability must be updated
according to a Bayes' rule,
\begin{equation} \label{Equation::Bayes}
    dp\big(B|d\,\Xi_{\,t},\Xi_{[0,t)}\big) =
    dq\big(d\,\Xi_{\,t}|B,\Xi_{[0,t)}\big)
    p\big(B|\Xi_{[0,t)}\big)
\end{equation}
where $dq\big(d\,\Xi_{\,t}|B,\Xi_{[0,t)}\big)$ is an infinitesimal
conditional probability that describes the likelihood of the
evolving measurement record, $d\,\Xi_{\,t}$, given a field with
magnitude $B$ and past history, $\Xi_{[0,t)}$. The utility of
Bayes' rule is that $q\big(d\,\Xi_{\,t}|B,\Xi_{[0,t)}\big)$ can be
computed using quantum trajectory theory, Eqs.\
(\ref{Equation::MeanJz}) and (\ref{Equation::DeltaJz2}).

Implementing this parameter estimator is best accomplished by a
(recursive) Kalman filter
\cite{Belavkin1999,Mabuchi1996,Verstraete2001}.  It can be shown
that the filtering equations,
\begin{equation} \label{Equation::Kalman}
    d \tilde{x}  = \mathbf{A} \tilde{x}\, dt
        + D^{-2} (\mathbf{B}+\mathbf{VC}^T)
        (d\,\Xi - \mathbf{C}\tilde{x}\,dt)
\end{equation}
with $\tilde{x}\equiv\left(\begin{array}{cc}\tilde{J}_\mathrm{z} &
\tilde{B}\\\end{array} \right)^T$, ($\tilde{J}_\mathrm{z}$ is the
estimate of $\langle \hat{J}_\mathrm{z} \rangle_c$),
$$
    \mathbf{A} = \gamma J e^\frac{-Mt}{2} \left( \begin{array}{cc}
      0 & 1 \\
      0 & 0 \\
    \end{array} \right)
    , \,\,\,
    \mathbf{B} = \left( \begin{array}{c} \langle \Delta
      \hat{J}_\mathrm{z}^2\rangle \\ 0 \end{array}\right)
    , \,\,\, \mathbf{C} = \left( \begin{array}{cc}
      1 & 0 \\ \end{array} \right) ,
$$
$D=1/(2\sqrt{M\eta})$, and $\tilde{x}(0) = 0$ implement 
Eq.\ (\ref{Equation::Bayes}).  We note that it is possible to
extend the Kalman filter to account for time-varying or stochastic
fields \cite{Stockton2003} as well as to implement quantum
feedback control \cite{Stockton2003,Thomsen2002}.

The conditional quantum dynamics, particularly spin-squeezing and
the exponential decay of the Bloch vector, enter the estimation
process via the covariance matrix,
\begin{equation} \label{Equation::VMatrixDef}
    \mathbf{V}(t) = \left( \begin{array}{cc}
      \Delta \tilde{J}_\mathrm{z}^2(t) &
      \Delta (\tilde{J}_\mathrm{z}\tilde{B})(t) \\
      \Delta (\tilde{B} \tilde{J}_\mathrm{z})(t) &
      \Delta \tilde{B}^2(t) \\
    \end{array} \right)
\end{equation}
which describes the uncertainty in the parameter estimations of
$\hat{J}_\mathrm{z}$ and $B$.  $\mathbf{V}(t)$ evolves
deterministically according to the matrix Riccati equation,
\begin{eqnarray}
    \dot\mathbf{V} & = & \left(\mathbf{A}-D^{-2}\mathbf{BC}\right)
    \mathbf{V} + \mathbf{V}\left(\mathbf{A}-D^{-2}\mathbf{BC}\right)^T
    \label{Equation::Riccati}\\
    & & \,+ \,D^{-2} \mathbf{V}
    \left(\mathbf{C}^T\mathbf{C}\right)\mathbf{V}
    \nonumber
\end{eqnarray}
subject to the initial conditions $\Delta
\tilde{J}_\mathrm{z}^2(0)= 0$ and
$\Delta(\tilde{J}_\mathrm{z}\tilde{B})(0) = 0$, with
$\Delta\tilde{B}^2(0)$ chosen to reflect prior knowledge on the
distribution of magnetic field values. Lacking any such knowledge,
one can set $\Delta\tilde{B}^2(0)\to\infty$.

The smallest detectable magnetic field as a function of $J$ and
the measurement duration, $t$, is determined by the estimator
variance, $\Delta \tilde{B}^2$. Solving the the matrix Riccati
equation [which is analytically soluble for the Kalman filter in
Eq.\ (\ref{Equation::Kalman})], provides the time-dependent
magnetic field detection threshold, $\delta \tilde{B}
\equiv\sqrt{\Delta\tilde{B}^2(t)}$,
\begin{equation} \label{Equation::QPELimit}
     \delta \tilde{B}(t) = \frac{M}{4\gamma J}
     \sqrt{\frac{ (1+ 2\eta J M t)}
     {a e^{-M t} + 4 e^{-M t /2}(4 \eta J + 1) + b} }
\end{equation}
with $a$ and $b$ given by,
\begin{eqnarray}
    a & = & -(2\eta J(M t + 4) + 1)\nonumber \\
    b & = & Mt + 2 \eta J (M t- 4)-3 \,\, . \nonumber
\end{eqnarray}
Expanding Eq.\ (\ref{Equation::QPELimit}) to leading order in $t$
provides an expression for the detection threshold,
\begin{equation} \label{Equation::ApproxLimit}
    \delta \tilde{B}(t) \approx \frac{1}{\gamma J}
    \sqrt{\frac{3}{M \eta t^3}}, \quad t \gg (J M)^{-1}.
\end{equation}
that is directly comparable to Eq.\
(\ref{Equation::ShotnoiseLimit}) when the measurement strength is
chosen to be $M \sim T_2^{-1}$ such that maximal spin-squeezing is
achieved at time $t=M^{-1}$.  Such a choice for $M$ permits a
superior $1/J$ (equivalently $1/N$) scaling that is characteristic
of the Heisenberg squeezing limit \cite{Kitagawa1993,Thomsen2002}.
The optical shotnoise [of order unity in this model, see Eq.\
(\ref{Equation::Photocurrent})] enters implicitly through the
signal to noise ratio, SNR=$J\sqrt{M}$, which highlights the
utility of the Kalman filter as a whitening filter--- it extracts
the non-stationary spin-squeezing and Larmor precession dynamics
despite the presence of Gaussian noise.

\begin{figure}[b*]
\vspace{-3mm}
\includegraphics{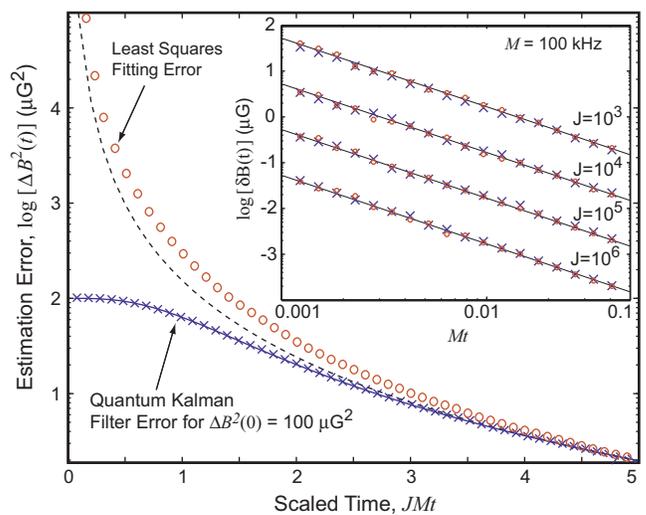}
\vspace{-3mm} \caption{Comparison of the estimation errors for a
quantum Kalman filter and a linear least squares magnetic field
determination.  The inset plot highlights the $1/J$ (Heisenberg
limited) scaling of both procedures for $t\gg (JM)^{-1}$.
\label{Figure::Data}}
\end{figure}

Figure \ref{Figure::Data} shows numerical results that demonstrate
the performance of our quantum Kalman filter (QKF).  The
simulations were performed for an atomic ensemble with $J=4 \times
10^6$, $\gamma = 1$ kHz/mG, $\eta=1$, and $M=100$ kHz in a
background magnetic field of $B=1$ $\mu$G.  These values nearly
correspond to a magnetometer constructed from $N=10^6$ ground
state Cs atoms coupled to a high-finesse optical cavity with
single photon Rabi frequency $g_0$ = 10 MHz and decay rate,
$\kappa$ = 1 MHz. The QND measurement corresponds to a
phase-quadrature homodyne detection of the transmitted cavity
light with a cavity mode ($P = 100$ $\mu$W) that is blue detuned
by $\Delta = 1$ GHz from the Cs transition at $\lambda \sim 852$
nm. The initial estimator variance was chosen to be $\Delta
\tilde{B}^2(0)=100$ $\mu$G$^2$ which is the initial value that one
would select given prior knowledge that the magnetic field could
be treated as a Gaussian random variable with this variance.  For
the parameters we selected, Larmor precession and spin projection
noise have comparable magnitudes on timescales of order
$(JM)^{-1}$.

The Kalman estimation error (crosses in Fig.\ \ref{Figure::Data})
was computed from the ensemble average, $E[(\tilde{B}_i-B)^2]$,
for $10^5$ trajectories, and the solid line shows the estimation
uncertainty $\Delta\tilde{B}(t)$ obtained by integrating Eq.\
(\ref{Equation::Riccati}). Since our simulations were performed
with $B=1$ $\mu$G, the empirical performance of the QKF closely
matches the solid line.  The dotted line in Fig.\
\ref{Figure::Data} shows the analytic Riccati solution, given by
Eq.\ (\ref{Equation::QPELimit}), for
$\Delta\tilde{B}^2(0)\rightarrow\infty$, which would be the
expected QKF performance in a scenario with no prior knowledge of
the magnetic field.  Fig.\ \ref{Figure::Data} also shows the
estimation error for simple linear regression of the measurement
record (open circles). Assuming that the Bloch vector has not
decayed significantly, $\tilde{B}$ is proportional to the slope of
a line fit to the (filtered) photocurrent, $y(t)$. The estimation
error for linear regression was obtained by computing
$E[(\tilde{B}_i-B)^2]$ for $10^5$ trajectories. Although the QKF
is clearly superior for short times, the two estimation procedures
converge for $(JM)^{-1} \ll t < M^{-1}$ and provide a quantum
parameter estimation with $\delta \tilde{B} \sim$ 0.01 nG in $t
\sim 1$ ms.  The inset of Fig.\ \ref{Figure::Data} highlights the
$\delta\tilde{B} \propto J^{-1}$ scaling that distinguishes both
the QKF (crosses) and regression (open circle) estimators from the
conventional shot-noise limit, Eq.\
\ref{Equation::ShotnoiseLimit}.  For sufficiently large times both
estimators achieve the detection threshold in Eq.\
\ref{Equation::ApproxLimit} (solid lines).

The QKF and linear regression differ mainly in how they treat the
initial diffusive transient of $\hat{J}_\mathrm{z}$ [Fig.\
\ref{Figure::Photocurrent}(B)]. Since the QKF is derived from a
quantum trajectory model, it is aware of the short-time diffusion
and strategically under-weights the photocurrent at early times
[via the Kalman gain, $\mathbf{G} \equiv D^{-2}(\mathbf{B} +
\mathbf{VC}^T)$]. At late times the regression analysis manages to
absorb the initial diffusive transient into the $y$-intercept of
the linear fit. Although $\langle \hat{J}_\mathrm{z}(0)\rangle =
0$, the $\hat{J}_\mathrm{z}$ localization process
\textit{gradually} determines an effective offset in the
photocurrent during the interval $t \sim (JM)^{-1}$. Without
explicit knowledge of the conditional dynamics the linear
regression equally weights the photocurrent for $t < (JM)^{-1}$.
This decreases the quality of the fit, but the resulting error
becomes insignificant for $t\gg (JM)^{-1}$.

Our analysis suggests that estimation procedures based on
conditional quantum dynamics can play a crucial role in optimizing
both the sensitivity and the bandwidth in atomic magnetometry.
While conventional steady-state magnetometers can only improve
their detection capabilities by increasing the number of atoms or
the averaging time, the quantum estimator can achieve greater
precision for the same value of $t$ and $N$ by improving the
measurement strength.  The significance of the Kalman filter is
the optimality that is guaranteed by its derivation from a Bayes'
rule.  Our finding that linear regression closely approximates the
optimal procedure indicates a potentially simpler experimental
procedure for sub-shotnoise magnetometry.   Although Heisenberg
limited spin-squeezing should be possible using current techniques
in cavity quantum electrodynamics (a discussion is provided in
\cite{Thomsen2002}), the experimental difficulty of achieving this
limit makes it desirable to have an optimal estimator such as the
QKF to fully exploit even a small amount of squeezing, to treat
fluctuating fields and to achieve estimator robustness
\cite{Stockton2003}. In either case, estimation procedures that
allow and account for conditional quantum dynamics--- whether
explicitly as in the QKF or implicitly as in linear regression---
offer substantial improvement over steady-state procedures.

This work was supported by the NSF (PHY-9987541, EIA-0086038), the
ONR (N00014-00-1-0479), and the Caltech MURI Center for Quantum
Networks (DAAD19-00- 1-0374). JKS acknowledges a Hertz fellowship.
Please visit http://minty.caltech.edu/Ensemble for simulation
source code and detailed notes on the Kalman filter derivation.


\begin{thebibliography}{17}
\expandafter\ifx\csname natexlab\endcsname\relax\def\natexlab#1{#1}\fi
\expandafter\ifx\csname bibnamefont\endcsname\relax
  \def\bibnamefont#1{#1}\fi
\expandafter\ifx\csname bibfnamefont\endcsname\relax
  \def\bibfnamefont#1{#1}\fi
\expandafter\ifx\csname citenamefont\endcsname\relax
  \def\citenamefont#1{#1}\fi
\expandafter\ifx\csname url\endcsname\relax
  \def\url#1{\texttt{#1}}\fi
\expandafter\ifx\csname urlprefix\endcsname\relax\def\urlprefix{URL }\fi
\providecommand{\bibinfo}[2]{#2}
\providecommand{\eprint}[2][]{\url{#2}}

\bibitem[{\citenamefont{Kominis et~al.}(2003)\citenamefont{Kominis, Kornack,
  Allred, and Romalis}}]{Romalis2003}
\bibinfo{author}{\bibfnamefont{I.}~\bibnamefont{Kominis}},
  \bibinfo{author}{\bibfnamefont{T.}~\bibnamefont{Kornack}},
  \bibinfo{author}{\bibfnamefont{J.}~\bibnamefont{Allred}}, \bibnamefont{and}
  \bibinfo{author}{\bibfnamefont{M.}~\bibnamefont{Romalis}},
  \bibinfo{journal}{Nature} \textbf{\bibinfo{volume}{422}},
  \bibinfo{pages}{596} (\bibinfo{year}{2003}).

\bibitem[{\citenamefont{Belavkin}(1999)}]{Belavkin1999}
\bibinfo{author}{\bibfnamefont{V.}~\bibnamefont{Belavkin}},
  \bibinfo{journal}{Rep. on Math. Phys.} \textbf{\bibinfo{volume}{43}},
  \bibinfo{pages}{405} (\bibinfo{year}{1999}).

\bibitem[{\citenamefont{Dupont-Roc et~al.}(1969)\citenamefont{Dupont-Roc,
  Haroche, and Cohen-Tannoudji}}]{CohenTannoudji1969}
\bibinfo{author}{\bibfnamefont{J.}~\bibnamefont{Dupont-Roc}},
  \bibinfo{author}{\bibfnamefont{S.}~\bibnamefont{Haroche}}, \bibnamefont{and}
  \bibinfo{author}{\bibfnamefont{C.}~\bibnamefont{Cohen-Tannoudji}},
  \bibinfo{journal}{Phys. Lett. A} \textbf{\bibinfo{volume}{28}},
  \bibinfo{pages}{638} (\bibinfo{year}{1969}).

\bibitem[{\citenamefont{Budker et~al.}(2002)\citenamefont{Budker, Gawlik,
  Kimball, Rochester, Yashchuk, and Weiss}}]{Budker2002}
\bibinfo{author}{\bibfnamefont{D.}~\bibnamefont{Budker}},
  \bibinfo{author}{\bibfnamefont{W.}~\bibnamefont{Gawlik}},
  \bibinfo{author}{\bibfnamefont{D.}~\bibnamefont{Kimball}},
  \bibinfo{author}{\bibfnamefont{S.}~\bibnamefont{Rochester}},
  \bibinfo{author}{\bibfnamefont{V.}~\bibnamefont{Yashchuk}}, \bibnamefont{and}
  \bibinfo{author}{\bibfnamefont{A.}~\bibnamefont{Weiss}},
  \bibinfo{journal}{Rev. Mod. Phys.} \textbf{\bibinfo{volume}{74}},
  \bibinfo{pages}{1153} (\bibinfo{year}{2002}).

\bibitem[{\citenamefont{Kim and Lee}(1998)}]{kim1998}
\bibinfo{author}{\bibfnamefont{C.}~\bibnamefont{Kim}} \bibnamefont{and}
  \bibinfo{author}{\bibfnamefont{H.}~\bibnamefont{Lee}}, \bibinfo{journal}{Rev.
  Sci. Inst.} \textbf{\bibinfo{volume}{69}}, \bibinfo{pages}{4152}
  (\bibinfo{year}{1998}).

\bibitem[{\citenamefont{Itano et~al.}(1993)\citenamefont{Itano, Berquist,
  Bollinger, Gilligan, Heinzen, Moore, Raizen, and
  D.J.Wineland}}]{Wineland1993}
\bibinfo{author}{\bibfnamefont{W.}~\bibnamefont{Itano}},
  \bibinfo{author}{\bibfnamefont{J.}~\bibnamefont{Berquist}},
  \bibinfo{author}{\bibfnamefont{J.}~\bibnamefont{Bollinger}},
  \bibinfo{author}{\bibfnamefont{J.}~\bibnamefont{Gilligan}},
  \bibinfo{author}{\bibfnamefont{D.}~\bibnamefont{Heinzen}},
  \bibinfo{author}{\bibfnamefont{F.}~\bibnamefont{Moore}},
  \bibinfo{author}{\bibfnamefont{M.}~\bibnamefont{Raizen}}, \bibnamefont{and}
  \bibinfo{author}{\bibnamefont{D.J.Wineland}}, \bibinfo{journal}{Phys. Rev. A}
  \textbf{\bibinfo{volume}{47}}, \bibinfo{pages}{3554} (\bibinfo{year}{1993}).

\bibitem[{\citenamefont{Kitagawa and Ueda}(1993)}]{Kitagawa1993}
\bibinfo{author}{\bibfnamefont{M.}~\bibnamefont{Kitagawa}} \bibnamefont{and}
  \bibinfo{author}{\bibfnamefont{M.}~\bibnamefont{Ueda}},
  \bibinfo{journal}{Phys. Rev. A} \textbf{\bibinfo{volume}{47}},
  \bibinfo{pages}{5138} (\bibinfo{year}{1993}).

\bibitem[{\citenamefont{Takahashi et~al.}(1999)\citenamefont{Takahashi, Honda,
  Tanaka, Toyoda, Ishikawa, and Yabuzaki}}]{Takahashi1999}
\bibinfo{author}{\bibfnamefont{Y.}~\bibnamefont{Takahashi}},
  \bibinfo{author}{\bibfnamefont{K.}~\bibnamefont{Honda}},
  \bibinfo{author}{\bibfnamefont{N.}~\bibnamefont{Tanaka}},
  \bibinfo{author}{\bibfnamefont{K.}~\bibnamefont{Toyoda}},
  \bibinfo{author}{\bibfnamefont{K.}~\bibnamefont{Ishikawa}}, \bibnamefont{and}
  \bibinfo{author}{\bibfnamefont{T.}~\bibnamefont{Yabuzaki}},
  \bibinfo{journal}{Phys. Rev. A} \textbf{\bibinfo{volume}{60}},
  \bibinfo{pages}{4974} (\bibinfo{year}{1999}).

\bibitem[{\citenamefont{Kuzmich et~al.}(2000)\citenamefont{Kuzmich, Mandel, and
  Bigelow}}]{Kuzmich2000}
\bibinfo{author}{\bibfnamefont{A.}~\bibnamefont{Kuzmich}},
  \bibinfo{author}{\bibfnamefont{L.}~\bibnamefont{Mandel}}, \bibnamefont{and}
  \bibinfo{author}{\bibfnamefont{N.~P.} \bibnamefont{Bigelow}},
  \bibinfo{journal}{Phys. Rev. Lett.} \textbf{\bibinfo{volume}{85}},
  \bibinfo{pages}{1594} (\bibinfo{year}{2000}).

\bibitem[{\citenamefont{Thomsen et~al.}(2002)\citenamefont{Thomsen, Mancini,
  and Wiseman}}]{Thomsen2002}
\bibinfo{author}{\bibfnamefont{L.}~\bibnamefont{Thomsen}},
  \bibinfo{author}{\bibfnamefont{H.}~\bibnamefont{Mancini}}, \bibnamefont{and}
  \bibinfo{author}{\bibfnamefont{H.}~\bibnamefont{Wiseman}},
  \bibinfo{journal}{Phys.\ Rev.\ A} \textbf{\bibinfo{volume}{65}},
  \bibinfo{pages}{061801} (\bibinfo{year}{2002}).

\bibitem[{\citenamefont{Carmichael}(1993)}]{Carmichael1993}
\bibinfo{author}{\bibfnamefont{H.}~\bibnamefont{Carmichael}},
  \emph{\bibinfo{title}{An open systems approach to quantum optics}}
  (\bibinfo{publisher}{Springer-Verlag}, \bibinfo{address}{New York},
  \bibinfo{year}{1993}).

\bibitem[{\citenamefont{Wiseman and Milburn}(1994)}]{Wiseman1994}
\bibinfo{author}{\bibfnamefont{H.}~\bibnamefont{Wiseman}} \bibnamefont{and}
  \bibinfo{author}{\bibfnamefont{G.}~\bibnamefont{Milburn}},
  \bibinfo{journal}{Phys. Rev. A} \textbf{\bibinfo{volume}{49}},
  \bibinfo{pages}{1350} (\bibinfo{year}{1994}).

\bibitem[{\citenamefont{Mabuchi}(1996)}]{Mabuchi1996}
\bibinfo{author}{\bibfnamefont{H.}~\bibnamefont{Mabuchi}},
  \bibinfo{journal}{Quantum Semiclass. Opt.} \textbf{\bibinfo{volume}{8}},
  \bibinfo{pages}{1103} (\bibinfo{year}{1996}).

\bibitem[{\citenamefont{Verstraete et~al.}(2001)\citenamefont{Verstraete,
  Doherty, and Mabuchi}}]{Verstraete2001}
\bibinfo{author}{\bibfnamefont{F.}~\bibnamefont{Verstraete}},
  \bibinfo{author}{\bibfnamefont{A.}~\bibnamefont{Doherty}}, \bibnamefont{and}
  \bibinfo{author}{\bibfnamefont{H.}~\bibnamefont{Mabuchi}},
  \bibinfo{journal}{Phys. Rev. A} \textbf{\bibinfo{volume}{64}},
  \bibinfo{pages}{032111} (\bibinfo{year}{2001}).

\bibitem[{\citenamefont{Stockton et~al.}(2003)\citenamefont{Stockton, Geremia,
  Doherty, and Mabuchi}}]{Stockton2003}
\bibinfo{author}{\bibfnamefont{J.~K.} \bibnamefont{Stockton}},
  \bibinfo{author}{\bibfnamefont{J.}~\bibnamefont{Geremia}},
  \bibinfo{author}{\bibfnamefont{A.~C.} \bibnamefont{Doherty}},
  \bibnamefont{and} \bibinfo{author}{\bibfnamefont{H.}~\bibnamefont{Mabuchi}},
  \bibinfo{journal}{quant-ph/0309101}  (\bibinfo{year}{2003}).

\bibitem[{\citenamefont{Silberfarb and Deutsch}(2003)}]{Deutsch2003}
\bibinfo{author}{\bibfnamefont{A.}~\bibnamefont{Silberfarb}} \bibnamefont{and}
  \bibinfo{author}{\bibfnamefont{I.}~\bibnamefont{Deutsch}},
  \bibinfo{journal}{Phys. Rev. A} \textbf{\bibinfo{volume}{68}},
  \bibinfo{pages}{013817} (\bibinfo{year}{2003}).

\bibitem[{\citenamefont{Smith et~al.}(2003)\citenamefont{Smith, Chaudhury, and
  Jessen}}]{Jessen2003}
\bibinfo{author}{\bibfnamefont{G.~A.} \bibnamefont{Smith}},
  \bibinfo{author}{\bibfnamefont{S.}~\bibnamefont{Chaudhury}},
  \bibnamefont{and} \bibinfo{author}{\bibfnamefont{P.~S.}
  \bibnamefont{Jessen}}, \bibinfo{journal}{J. Opt. B: Quant. Semiclass. Opt.}
  \textbf{\bibinfo{volume}{5}}, \bibinfo{pages}{323} (\bibinfo{year}{2003}).

\end{thebibliography}


\vspace{-6mm}
\end{document}